\documentclass[pdflatex,sn-mathphys]{sn-jnl}

\jyear{2022}%

\theoremstyle{thmstyleone}%
\theoremstyle{thmstyletwo}%

\theoremstyle{thmstylethree}%

\begin{document}

\title[New opportunities for kaonic atoms measurements from CdZnTe detectors]{New opportunities for kaonic atoms measurements from CdZnTe detectors}


\author[1]{\fnm{L.} \sur{Abbene}}
\author[2]{\fnm{M.} \sur{Bettelli}}
\author[1]{\fnm{A.} \sur{Buttacavoli}}
\author[1]{\fnm{F.} \sur{Principato}}
\author[2]{\fnm{A.} \sur{Zappettini}}
\author[3]{\fnm{C.} \sur{Amsler}}
\author[4]{\fnm{M.} \sur{Bazzi}}
\author[5]{\fnm{D.} \sur{Bosnar}}
\author[6]{\fnm{M.} \sur{Bragadireanu}}
\author[3]{\fnm{M.} \sur{Cargnelli}}
\author[7]{\fnm{M.} \sur{Carminati}}
\author[4]{\fnm{A.} \sur{Clozza}}
\author[7]{\fnm{G.} \sur{Deda}}
\author[4]{\fnm{L.} \sur{De Paolis}}
\author[8,4]{\fnm{R.} \sur{Del Grande}}
\author[8]{\fnm{L.} \sur{Fabbietti}}
\author[7]{\fnm{C.} \sur{Fiorini}}
\author[5]{\fnm{I.} \sur{Fri\v{s}\v{c}i\'{c}}}
\author[4]{\fnm{C.} \sur{Guaraldo}}
\author[4]{\fnm{M.} \sur{Iliescu}}
\author[9]{\fnm{M.} \sur{Iwasaki}}
\author[4]{\fnm{A.} \sur{Khreptak}}
\author[4]{\fnm{S.} \sur{Manti}}
\author[3]{\fnm{J.} \sur{Marton}}
\author[4]{\fnm{M.} \sur{Miliucci}}
\author[10,11]{\fnm{P.} \sur{Moskal}}
\author[4]{\fnm{F.} \sur{Napolitano}}
\author[10,11]{\fnm{S.} \sur{Nied\'zwiecki}}
\author[12]{\fnm{H.} \sur{Ohnishi}}
\author[13,4]{\fnm{K.} \sur{Piscicchia}}
\author[12]{\fnm{Y.} \sur{Sada}}
\author[4]{\fnm{F.} \sur{Sgaramella}}
\author[3]{\fnm{H.} \sur{Shi}}
\author[10,11]{\fnm{M.} \sur{Silarski}}
\author[4,13,6]{\fnm{D. L.} \sur{Sirghi}}
\author[4,6]{\fnm{F.} \sur{Sirghi}}
\author[10,11]{\fnm{M.} \sur{Skurzok}}
\author[4]{\fnm{A.} \sur{Spallone}}
\author[12]{\fnm{K.} \sur{Toho}}
\author[3,14]{\fnm{M.} \sur{T\"uchler}}
\author[4]{\fnm{O.} \sur{Vazquez Doce}}
\author[12]{\fnm{C.} \sur{Yoshida}}
\author[3]{\fnm{J.} \sur{Zmeskal}}
\author*[4]{\fnm{A.} \sur{Scordo}}\email{alessandro.scordo@lnf.infn.it}
\author[4]{\fnm{C.} \sur{Curceanu}}

\affil[1]{\orgdiv{Dipartimento di Fisica e Chimica - Emilio Segrè}, \orgname{Università di Palermo}, \orgaddress{\street{Viale Delle Scienze, Edificio 18}, \city{Palermo}, \postcode{90128}, \country{Italy}}}
\affil[2]{\orgdiv{Istituto Materiali per l’Elettronica e il Magnetismo}, \orgname{Consiglio Nazionale delle Ricerche}, \orgaddress{\street{Parco Area delle Scienze 37/A}, \city{Parma}, \postcode{43124}, \country{Italy}}}
\affil[3]{\orgname{Stefan-Meyer-Institut f\"ur subatomare Physik}, \orgaddress{\street{Kegelgasse 27}, \city{Vienna}, \postcode{1030}, \country{Austria}}}
\affil*[4]{\orgdiv{Laboratori Nazionali di Frascati}, \orgname{INFN}, \orgaddress{\street{Via E. Fermi 54}, \city{Frascati}, \postcode{00044}, \country{Italy}}}
\affil[5]{\orgdiv{Department of Physics, Faculty of Science }, \orgname{University of Zagreb}, \orgaddress{\street{Bijeni\v cka cesta 32}, \city{Zagreb}, \postcode{10000}, \country{Croatia}}}
\affil[6]{\orgname{Horia Hulubei National Institute of Physics and Nuclear Engineering (IFIN-HH)}, \orgaddress{\street{No. 30, Reactorului Street}, \city{Magurele, Ilfov}, \postcode{077125}, \country{Romania}}}
\affil[7]{\orgdiv{Dipartimento di Elettronica, Informazione e Bioingegneria and INFN Sezione di Milano}, \orgname{Politecnico di Milano}, \orgaddress{\street{Via Giuseppe Ponzio 34}, \city{Milano}, \postcode{20133}, \country{Italy}}}
\affil[8]{\orgdiv{Physik Department E62}, \orgname{Technische Universit\"at M\"unchen}, \orgaddress{\street{James-Franck-Stra{\ss}e 1}, \city{Garching}, \postcode{85748}, \country{Germany}}}
\affil[9]{\orgdiv{Institute of Physical and Chemical Research}, \orgname{RIKEN}, \orgaddress{\street{2-1 Hirosawa}, \city{Wako, Saitama}, \postcode{351-0198}, \country{Japan}}}
\affil[10]{\orgdiv{Faculty of Physics, Astronomy, and Applied Computer Science}, \orgname{Jagiellonian University}, \orgaddress{\street{ul. prof. Stanis\l awa \L ojasiewicza 11}, \city{Krak\'ow}, \postcode{30-348}, \country{Poland}}}
\affil[11]{\orgdiv{Center for Theranostics}, \orgname{Jagiellonian University}, \orgaddress{\street{ul. prof. Stanis\l awa \L ojasiewicza 11}, \city{Krak\'ow}, \postcode{30-348}, \country{Poland}}}
\affil[12]{\orgdiv{Research Center for Electron Photon Science (ELPH)}, \orgname{Tohoku University}, \orgaddress{\street{1-2-1 Mikamine}, \city{Taihaku-ku, Sendai, Miyagi}, \postcode{982-0826}, \country{Japan}}}
\affil[13]{\orgname{Centro Ricerche Enrico Fermi-Museo Storico della fisica e Centro Studi e Ricerche “Enrico Fermi”}, \orgaddress{\street{Via Panisperna, 89a}, \city{Rome}, \postcode{00184}, \country{Italy}}}
\affil[14]{\orgdiv{Vienna Doctoral School in Physics}, \orgname{University of Vienna}, \orgaddress{\street{Boltzmanngasse 5}, \city{Vienna}, \postcode{1090}, \country{Austria}}}


\abstract{We present the tests performed by the SIDDHARTA-2 collaboration at the $\mathrm{DA\Phi NE}$ collider with a quasi-hemispherical CdZnTe detector. 
The very good room-temperature energy resolution and efficiency in a wide energy range show that this detector technology is ideal for studying radiative transitions 
in intermediate and heavy mass kaonic atoms. The CdZnTe detector was installed for the first time in an accelerator environment
to perform tests on the background rejection capabilities, which were achieved by exploiting the SIDDHARTA-2 Luminosity Monitor.
A spectrum with an $\mathrm{^{241}Am}$ source has been acquired, with beams circulating in the main rings, 
and peak resolutions of 6\% at 60 keV and of 2.2\% at 511 keV have been achieved. 
The background suppression factor, which turned out to be of the order of $\mathrm{\simeq10^{5-6}}$, opens the possibility to plan for future kaonic atom measurements with CdZnTe detectors.}




\maketitle

\section{Introduction}\label{sec_intro}

\noindent Kaonic atoms are formed when a $\mathrm{K^-}$ is moderated inside a target until it reaches a low enough kinetic energy to be stopped, 
replacing one of the outer electrons and forming an exotic atom in a highly excited state. The kaonic atom then undergoes atomic cascade to the ground state.
These systems provide an ideal tool to study the low-energy regime of Quantum Chromodynamics (QCD) since, due to the much heavier $\mathrm{K^-}$ mass with respect to the 
$\mathrm{e^-}$ one, the lower levels are close enough to the nucleus to be influenced by the short-range strong interaction between the nucleus and 
the $\mathrm{K^-}$ \cite{Napolitano:2022eik}.\\
Kaonic atoms have been intensively studied in the 1970s and 1980s with a series of measurements, 
still representing today the main database for low-energy antikaon-nucleon studies \cite{Davies:1979,Izycki:1980,Bird:1983,Wiegand:1971zz,Baird:1983ub,Friedman:1994hx}.
More recently, a second generation of experiments performed new and more accurate measurements on kaonic hydrogen and helium \cite{Iwasaki:1997,Okada:2007ky,SIDDHARTA:2011dsy,SIDDHARTA:2009qht,SIDDHARTA:2010uae,SIDDHARTA:2012rsv},
whose results were in better agreement with the theoretical predictions but not compatible with the old experiments \cite{RevModPhys.91.025006}. \\
The technological boost in radiation detectors in the last decade stimulated the strangeness nuclear physics community to plan for new experiments aiming to renew the database on
kaonic atoms, in particular at the $\mathrm{DA\Phi NE}$ collider at the INFN National Laboratories of Frascati \cite{Zobov:2010zza}.
To detect the radiation emitted from the various transitions, a special role could be played by cadmium–zinc–telluride (CdZnTe) detectors, 
thanks to their excellent performances at room temperatures in terms of energy resolution
(few \% FWHM), linearity and fast readout (few tens of ns) in a broad energy region ranging from few keV to some MeV \cite{s90503491,Owens:gf0007}.\\
One of the most challenging difficulties to be overcome in a facility like $\mathrm{DA\Phi NE}$ is the high electromagnetic and hadronic background to which these detectors will be exposed, since their small active area requires to place them as close as possible to the Interaction Point (IP), to maximize the solid angle acceptance.\\
In this work we present the first tests ever performed with such types of detectors in an accelerator; they were specifically aimed to assess the performances in the presence of a high machine background, similarly to what was already performed for the Silicon Drift Detectors by the SIDDHARTA-2 collaboration \cite{https://doi.org/10.48550/arxiv.2208.03422}.
Their fast timing response allows for a high-effective background suppression, achieved by using a trigger system based on kaons detection via Time of Flight (ToF), crucial for the possibility to observe the radiation emitted from the kaonic atom transitions \cite{Skurzok:2020phi}.

\section{Experimental setup}\label{sec_setup}

\begin{figure}[h]%
\centering
	\includegraphics[width=0.9\textwidth]{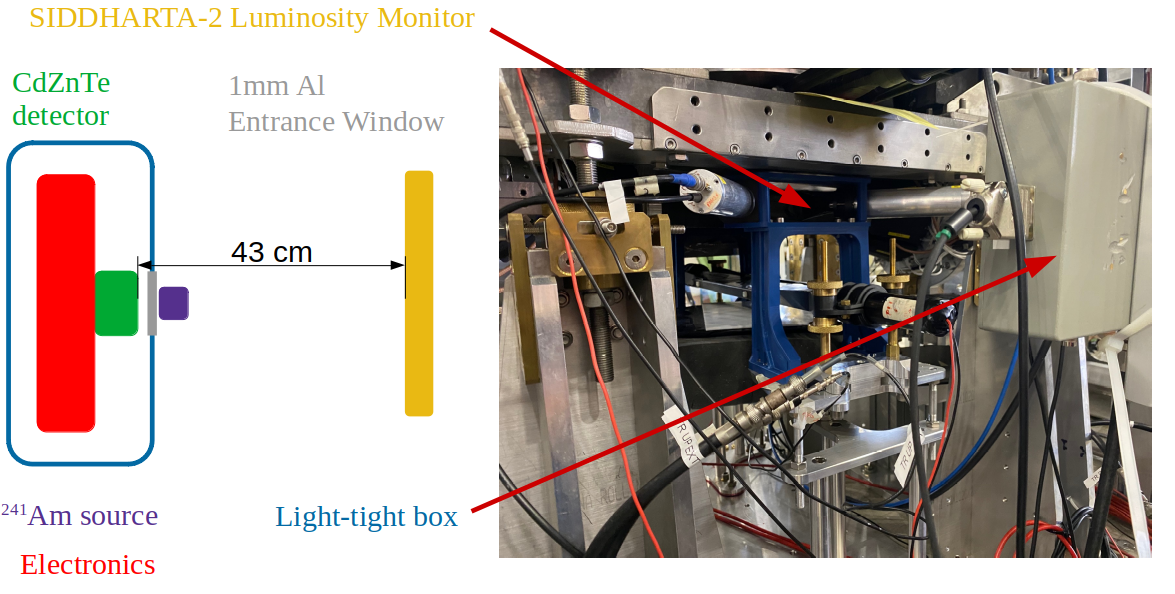}
	\caption{\emph{Left:} Top view of the geometry of the setup. \emph{Right:} Picture of the light-tight box containing the CdZnTe prototype installed at the $\mathrm{DA\Phi NE}$ collider.}\label{fig:setup}
\end{figure}

In June 2022, a first prototype of a quasi-hemispherical CdZnTe detector system, with an active surface of $\mathrm{1\,cm^2}$ and a thickness of 5 mm, was installed near the IP of the $\mathrm{DA\Phi NE}$ collider, vertically aligned with the SIDDHARTA-2 Luminosity Monitor (LM) \cite{Skurzok:2020phi},
at a distance of 43 cm. A (not in scale) top view of the geometry of the setup and a picture of the prototype as installed at $\mathrm{DA\Phi NE}$ are shown in Fig. \ref{fig:setup}.

\noindent The prototype, manufactured by the IMEM-CNR of Parma, was enclosed in a light-tight box with a 1 mm thick aluminum entrance window, matching the detector's active surface, on top of which an $\mathrm{^{241}Am}$ radioactive source was placed, producing a $\mathrm{\simeq500\,Hz}$ signal in the CdZnTe detector.\\
The detector was connected to low-noise (equivalent noise charge ENC of 100 electrons) charge-sensitive preamplifiers (CSPs) and processed by 8-channel digital electronics. Both the CSPs and the digital electronics were developed at DiFC of University of Palermo (Italy).
The signals from the CdZnTe were acquired by two CAEN DT5724 digitizers driven by an original firmware \cite{Abbene:pp5105,GERARDI201446,Abbene_2013}.\\ 
The signals from the LM were also acquired, with the same digitizers, to perform an offline data selection (trigger) when a charged kaon pair was produced in the horizontal plane. 
The digital signals from the LM were processed by an ORTEC 566 Time-to-Amplitude Converter (TAC) module. 
The LM is based on the J-PET technology developed for the imaging of
electron-positron annihilation \cite{doi:10.1126/sciadv.abh4394,Moskal:2021kxe}.
A detailed description of the LM and of the discrimination between the kaons and the Minimum Ionizing Particles (MIPs) produced during the 
$\mathrm{e^+e^-}$ collisions can be found in \cite{Skurzok:2020phi}.\\
The data reported in this paper have been acquired for a total of 72 hours, during which the accelerator delivered $\mathrm{e^-}$ and $\mathrm{e^+}$ beams 
with average currents of $\mathrm{\simeq500\,mA}$ and $\mathrm{\simeq270\,mA}$, respectively. 

\section{Results}\label{sec_results}

\begin{figure}[h]%
\centering
	\includegraphics[width=0.9\textwidth]{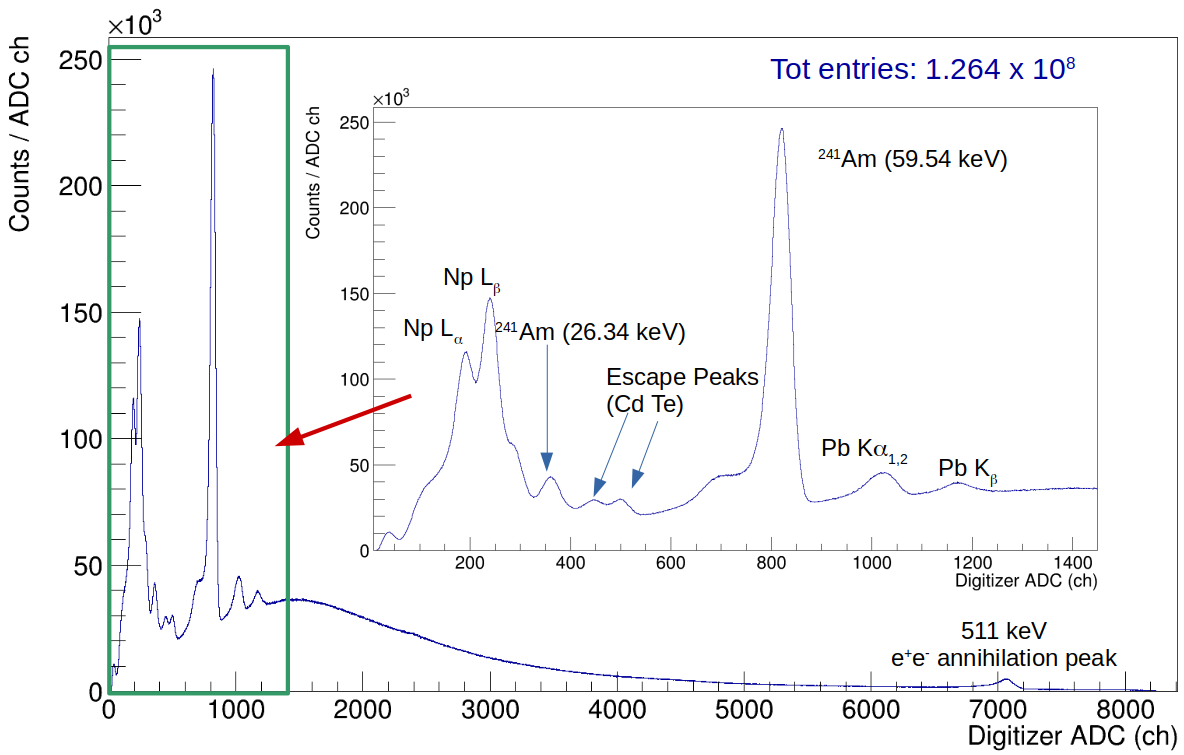}
	\caption{CdZnTe spectrum acquired in 72 hours at $\mathrm{DA\Phi NE}$ without any trigger requirement.}\label{fig:spectrum}
\end{figure}

The spectrum acquired in the 72 hours test with the CdZnTe detection system is plotted, without any data selection requirement, in Fig. \ref{fig:spectrum}. 

\noindent The spectrum was obtained without any selection. The $\mathrm{^{241}Am}$ peaks are visible together with the Neptunium (Np) ones, 
both coming from the source decay chain. The Lead (Pb) X-rays come from the activation of the SIDDHARTA-2 setup lead shielding by the MIPs lost from the beams.
Two escape peaks of the main $\mathrm{^{241}Am}$ peak, due to Cd and Te, are also observed \cite{Abbene:pp5105}.
The measured peak resolutions are of 6\% at 60 keV and of 2.2\% at 511 keV.\\
To simulate the rejection factor of the background of the $\mathrm{DA\Phi NE}$ machine, the TAC signals have been used to select only events in which a $\mathrm{K^+K^-}$ pair passed through the two scintillators of the LM.
In this way, the case in which only events correlated in time with the formation of kaonic atoms in a target placed in front of the CdZnTe entrance window 
are retained, is simulated.

\begin{figure}[h]%
\centering
	\includegraphics[width=0.9\textwidth]{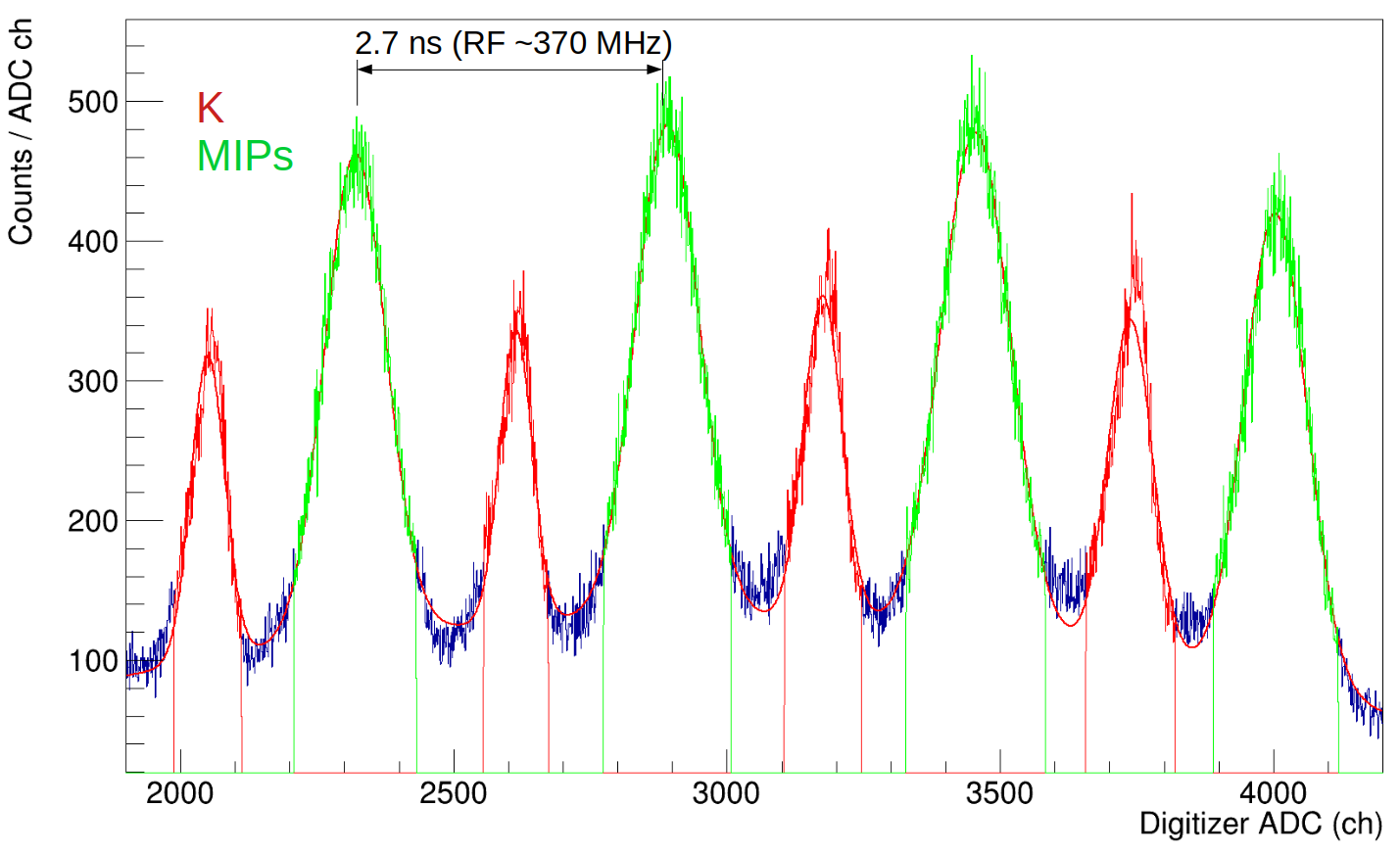}
	\includegraphics[width=0.9\textwidth]{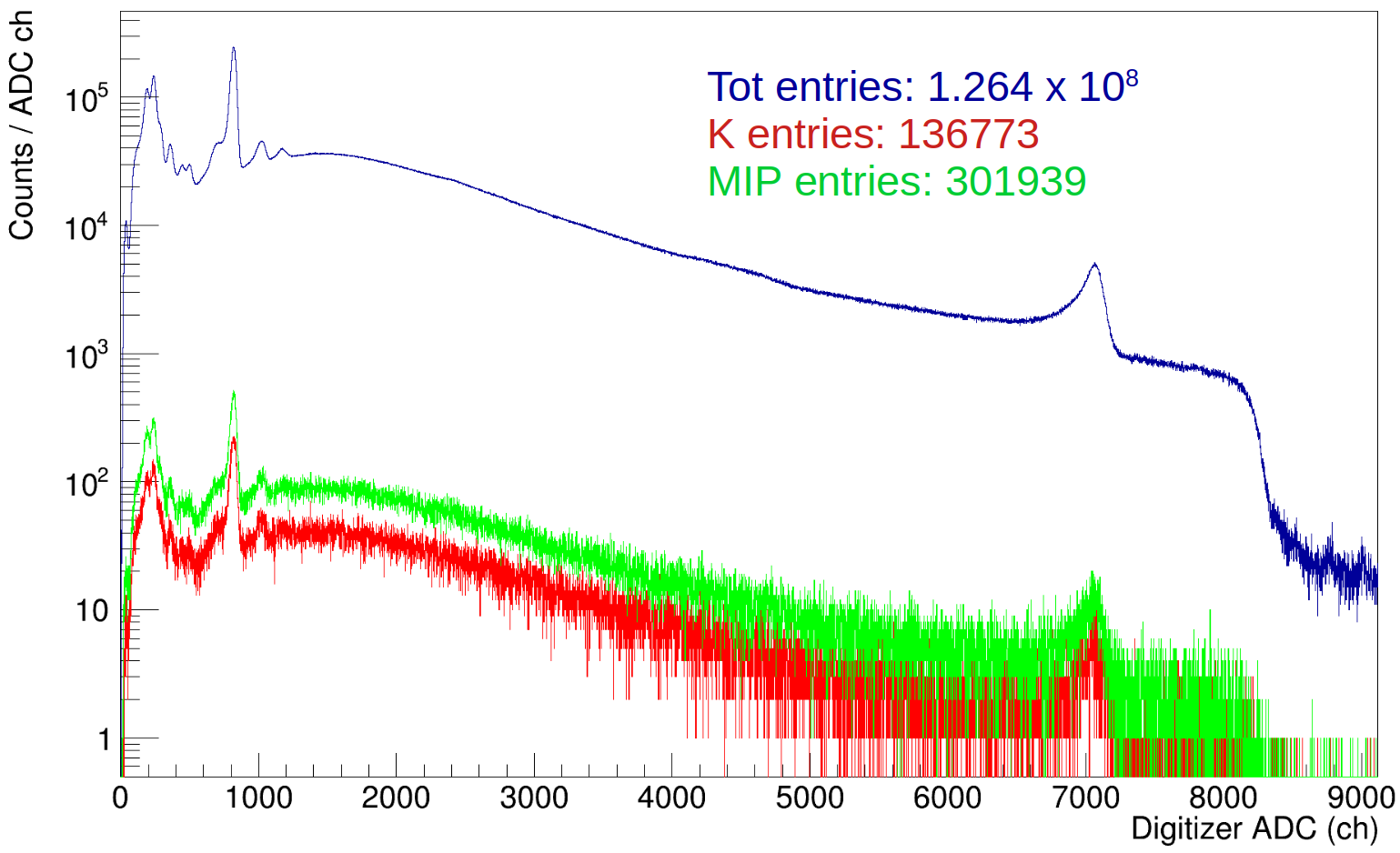}
	\caption{\emph{Top:} TAC spectrum from the SIDDHARTA-2 LM; red and green peaks correspond to kaons and MIPs, respectively. \emph{Bottom:} CdZnTe spectrum with no time selection from the TAC (blue) overimposed on the
spectra obtained in coincidence with kaons (red) or MIPs (green) events.}\label{fig:rejection}
\end{figure}

\begin{figure}[h]%
\centering
	\includegraphics[width=0.9\textwidth]{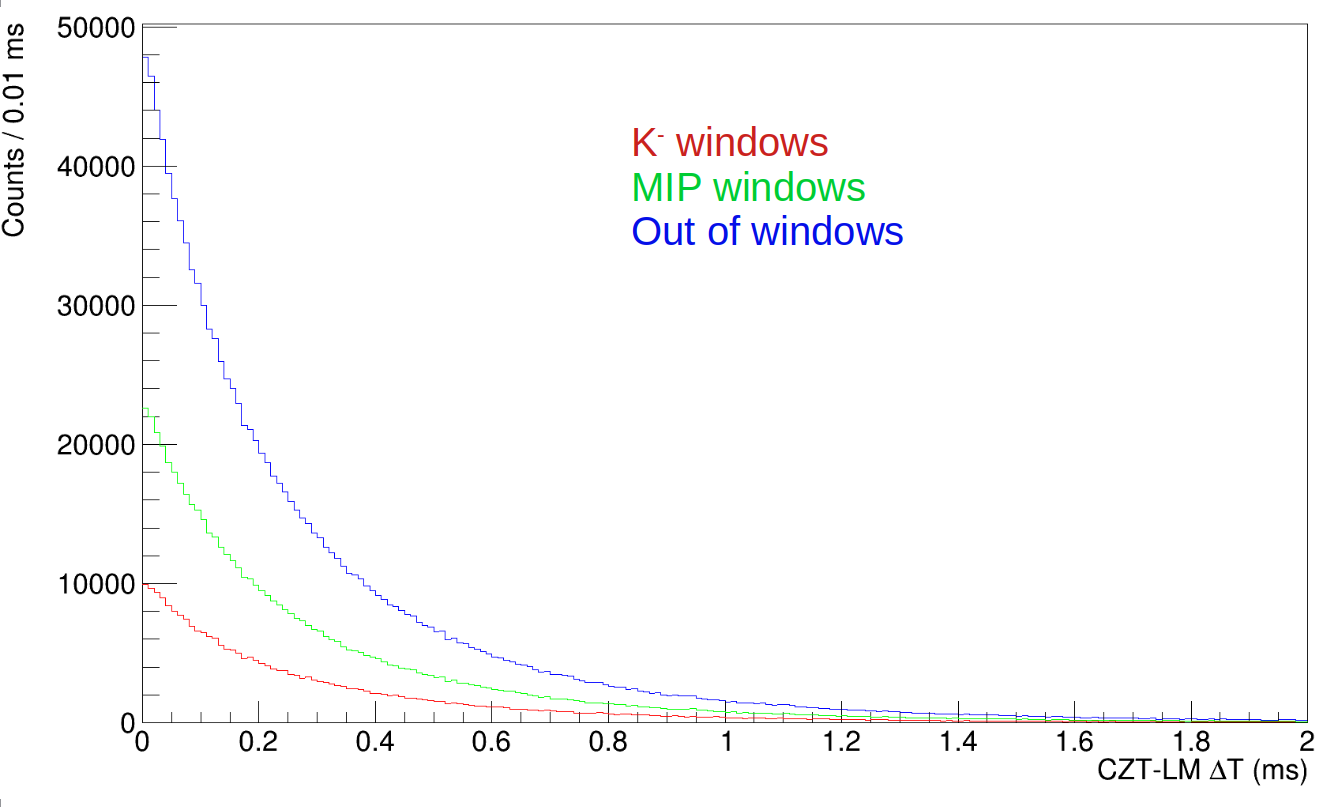}
	\caption{Distribution of the time difference between a photon detection by the CdZnTe detector and a signal in the SIDDHARTA-2 LM.}
	\label{fig:deltaT}
\end{figure}

\noindent In the upper part of Fig. \ref{fig:rejection}, the TAC spectrum from the SIDDHARTA-2 LM is shown, where red and green peaks correspond to kaons and MIPs, respectively.\\ 
The two LM scintillators are placed at a specific distance, with respect to the IP, such as to have a difference in the arrival time between kaons and MIPs producing two clear
peaks, visible in the figure.
The $\mathrm{\sim\,370\,MHz}$ radiofrequency (RF) of the $\mathrm{DA\Phi NE}$ collider is used in coincidence with the signals from the LM to get a few ps resolution timestamp
for each event. To provide a reference for each collision, a Constant Fraction Discriminator (CFD) must be employed. The CFDs are usually limited to work at 200 MHz;
to overcome this limitation, the RF/4 is then used, at a frequency of $\mathrm{\sim\,90\,MHz}$. As a consequence, every coincidence event in the 
LM discriminators can be randomly associated in time with one of the four collisions; this is reflected as the four double structures of Fig. \ref{fig:rejection}.\\
The events from the radioactive source are uncorrelated in time with the $\mathrm{DA\Phi NE}$ collisions; to test the background suppression, 
we thus require that a signal from the LM occurs just before a signal in the CdZnTe detector, simulating the case where X-rays from the transitions of kaonic atoms 
hit the detector immediately after the detection of a kaon pair in the LM. \\
The lower part of Fig. \ref{fig:rejection} shows the spectrum with no time selection from the TAC (blue) overimposed with the spectra obtained requiring the presence either of a kaon (red) or a MIP (green) previous to the CdZnTe signal.
The shapes of these spectra are all identical and reflect the absence, in this test, of a target where kaonic atoms can be formed. 
The trigger-induced suppression consists of the rejection of all those events in which a detected decay from the $\mathrm{^{241}Am}$ source is not (randomly) occurring 
in coincidence with the signal on the LM. 
The requirement of a kaon or a MIP signal within two CdZnTe events induces a suppression of the background of a factor $\mathrm{\simeq3\times10^2}$. \\
The events from the source occur randomly in time with respect to the collisions and, consequently, to the LM signals. The distributions of the time difference, $\mathrm{\Delta T}$, 
between a photon detection by the CdZnTe detector and a signal in the SIDDHARTA-2 LM are shown in Fig. \ref{fig:deltaT} for kaons (red), MIPs (green) and all the other events out of the K/MIP selections (blue). For each event on the LM, processed with the TAC and acquired with the digitizer, the distributions show the $\mathrm{\Delta T s}$ with respect to 
the subsequent event on the CdZnTe detector.

\noindent From Fig. \ref{fig:deltaT} it turns out that $\mathrm{^{241}Am}$ events occur with a time difference up to 2 ms, which is exactly what expected from the $\mathrm{500\,Hz}$
rate of the $\mathrm{^{241}Am}$ source.
When operated with a target to perform kaonic atoms measurements, the $\mathrm{\Delta T}$ distribution is expected to have two components: a flat background, collecting all
the non-kaonic events on the CdZnTe occurring randomly in time with respect to the LM ones, and a second one where, on the contrary, the events correlated with a $\mathrm{K^-}$
passing through the LM, stopping in the target and forming the kaonic atoms will produce a peak. \\
In a similar way to what was observed for the Silicon Drift Detectors (SDDs) of the 
SIDDHARTA-2 setup \cite{https://doi.org/10.48550/arxiv.2208.03422}, this peak will have a FWHM reflecting both the timescale of the physical processes described above and
that of the CdZnTe readout. In \cite{https://doi.org/10.48550/arxiv.2208.03422}, this FWHM is ranging from 430 to 950 ns, depending on the temperature of the SDDs,
but the faster readout of CdZnTe detectors suggests that smaller values could be achieved.
In Tab. \ref{tab:rej}, we report the number of events on the CdZnTe satisfying the request to have a coincidence with a $\mathrm{K^-}$ in the LM ($\mathrm{K^-_{TAC}}$) 
without any $\mathrm{\Delta T}$ request as well as with additional requirements of $\mathrm{\Delta T<1000,500,300,100\, ns}$. The resulting background suppression factors are also reported.

\renewcommand{\arraystretch}{2}
\begin{table}[h]
\centering
  \begin{tabular}{ | c | c | c | }
    \hline
    Request & Events & Rejection factor \\ \hline
    No request & $1.26\times10^8$ & 1 \\ \hline
	  $\mathrm{K^-_{TAC}}$ & 136359 & $3\times10^2$ \\ \hline
	  $\mathrm{K^-_{TAC},\Delta T < 1\,\mu s}$ & 1096 & $1.15\times10^{5}$ \\ \hline
	  $\mathrm{K^-_{TAC},\Delta T < 500\,ns}$ & 605 & $2.08\times10^{5}$ \\ \hline
	  $\mathrm{K^-_{TAC},\Delta T < 300\,ns}$ & 374 & $3.33\times10^{5}$  \\ \hline
	  $\mathrm{K^-_{TAC},\Delta T < 100\,ns}$ & 124 & $1.02\times10^{6}$  \\ \hline
  \end{tabular}\vspace{5mm}
	\caption{Number of events on the CdZnTe satisfying the request to have a coincidence with a $\mathrm{K^-}$ in the LM (TAC), without any $\mathrm{\Delta T}$ request 
	as well as with additional requirements of $\mathrm{\Delta T<1000,500,300,100\, ns}$, and the resulting background suppression factors.}
  \label{tab:rej}
\end{table}

\noindent Depending on the time window, suppression factors of the order of $\mathrm{\simeq10^{5-6}}$ are found. In particular, the request to 
have a TAC signal in time between two CdZnTe ones is responsible for the major suppression of the asynchronous background (mainly due to MIPs lost from the beams), while
further reductions of the coincidence time window not only contribute to this suppression, but also, requiring a $\mathrm{K^+K^-}$ pair in the LM, help in the 
rejection of the synchronous component of the background \cite{https://doi.org/10.48550/arxiv.2208.03422}.

\section{Conclusions}

In this work we presented the first tests performed with a quasi-hemispherical CdZnTe detector in an accelerator environment like the $\mathrm{DA\Phi NE}$ collider at the
INFN Laboratories of Frascati, where an $\mathrm{^{241}Am}$ spectrum was acquired for 72 hours, with the beams circulating in the main rings, showing peak resolutions 
of 6\% at 60 keV and of 2.2\% at 511 keV.\
We successfully measured ad suppression of the machine background with a trigger system, exploiting the fast readout capability of the device.
The $\mathrm{\simeq10^{5-6}}$ obtained rejection factors represent a very promising result which encourage the SIDDHARTA-2 collaboration to include, 
in its future data-taking, measurements of radiative transitions from several intermediate and high mass kaonic atoms to be performed with CdZnTe detectors.

\backmatter

\bmhead{Acknowledgments}

\noindent We thank C. Capoccia from LNF-INFN and H. Schneider, L. Stohwasser, and D. Pristauz- Telsnigg from Stefan-
Meyer-Institut for their fundamental contribution in designing and building the SIDDHARTA-2 setup. We thank as
well the $\mathrm{DA\Phi NE}$ staff for the excellent working conditions and permanent support. Part of this work was supported by 
the EU STRONG-2020 (JRA8) project within HORIZON 2020, Grant agreement ID: 824093; the Austrian Science Fund (FWF): [P24756-N20 and P33037-N]; 
the Croatian Science Foundation under the project IP-2018-01-8570; Japan Society for the Promotion of Science JSPS KAKENHI Grant No. JP18H05402; 
the Polish Ministry of Science and Higher Education Grant No. 7150/E-338/M/2018 and the Polish
National Agency for Academic Exchange (Grant no PPN/BIT/2021/1/00037); the SciMat and qLife Priority Research Areas budget under 
the program Excellence Initiative - Research University at the Jagiellonian University.

\section{*Declarations}

Data Availability Statement: The datasets generated during and/or analysed during the current study are available from the corresponding author on reasonable request.

\bibliography{epj_st_scordo}


\end{document}